# Calibration of the Milagro Cosmic Ray telescope.

**Lazar Fleysher**[1] **for the Milagro Collaboration**
[1]*Department of Physics, New York University, New York, NY 10003, USA*

### Abstract

The Milagro detector is an air shower array which uses the water Cherenkov technique and is capable of continuously monitoring the sky at energies near 1 TeV. The detector consists of 20000 metric tons of pure water instrumented with 723 photo-multiplier tubes (PMTs). The PMTs are arranged in a two-layer structure on a lattice of 3 m spacing covering 5000 $m^2$ area. The direction of the shower is determined from the relative timing of the PMT signals, necessitating a common time reference and amplitude slewing corrections to improve the time resolution. The calibration system to provide these consists of a pulsed laser driving 30 diffusing light sources deployed in the pond to allow cross-calibration of the PMTs. The system is capable of calibrating times and the pulse-heights from the PMTs using the time-over-threshold technique. The absolute energy scale is provided using single muons passing through the detector. The description of the calibration system of the Milagro detector and its prototype Milagrito will be presented.

## 1 Introduction

This paper describes the method which was used to calibrate the prototype called Milagrito and the Milagro detectors. Milagrito physics results are reported elsewhere in these procedeengs. The layout of Milagro detector is described in (McCullough et al., 1999), but, for clarity, some information is provided here.

Milagro is the first detector designed to study air showers at energies near 1 TeV using water Cherenkov techniques. The detector is built in the Jemez Mountains near Los Alamos, New Mexico at an altitude of 2650 m. The pond, which is 60m x 80m x 8m, is filled with clean water, covered with a light barrier and instrumented with 723 - 20 cm PMT's.

The PMTs collect Cherenkov light produced by the shower particles which traverse the detector's water volume. Whenever a PMT pulse exceeds a preset discriminator threshold a multihit time-to-digital converter (TDC) is started. Each PMT has its own TDC which is capable of recording up to 16 discriminator level crossings per event with 0.5 ns resolution. These constitute the raw data from the PMT.

The calibration procedure described below is applied in order to transform the raw counts to physically meaningful arrival times and light intensities which then can be used for event reconstruction. Considerable effort has been made to construct all PMT channels of the detector as uniformly as possible. However to achieve the high precision required for the event reconstruction, the remaining variations between channels have to be compensated for. A separate set of calibration parameters is determined for each PMT channel.

The calibration system has been designed to reflect the physics goals of the detector and is capable of calibrating the times recorded by each PMT to provide the best available shower direction; it is capable of calibrating the "pulse-heights" from the PMTs needed to estimate the absolute energy of each event. The absolute time of the events is retrieved from the GPS (Global Positioning System) clock system to 100 ns accuracy.

**1.1 Pointing** The desire to reconstruct the position of events on the Celestial sphere with systematic errors $\ll 1°$ dictates that the locations of the photo-tubes be known to about 10 cm accuracy in horizontal direction and about 3 cm in vertical. To meet this requirement, photographic and theodolite surveys of the pond have been performed. At the end of the construction period, when the pond was filled with water, an "as-built" measurement of the elevation of all PMTs has been made.

Knowledge of the PMT coordinates is necessary, although not sufficient, to achieve the stated goal. Times, registered by PMTs, have to have resolution of about 1 ns. This is limited by the transit time jitter of the PMTs at low light levels. Thus, it is important to calibrate the TDC conversion factors, compensate for the

PMT-pulse amplitude dependence of TDC measurements (known as a slewing correction) and synchronize all TDCs (find TDC time offsets) to the required accuracy.

**1.2 Energy** The relative "pulse-height"-to-photo-electron conversion must be determined to convert all amplitude measurements to a common unit for each event. The photo-electron counts (PEs) then has to be converted to the absolute scale of the energy deposited in the water to reconstruct the shower size and, ultimately, to estimate the energy of the primary particle.

## 2 Time-over-Threshold

Traditionally, the area of the PMT pulses is measured using amplitude-to-digital converters (ADCs). The major draw-back of this method is that ADCs have narrow dynamic range and are relatively slow devices which causes dead time during data taking. A new technique has been developed to overcome these problems.

The idea behind the Time-over-Threshold (ToT) method is simple. The PMT pulse quickly charges a capacitor $C$, which is then slowly discharged via a load resistor $R$. In such a setup the total area of a PMT pulse can be measured by the discharge time (time over threshold).

This method will work only if the time between registered pulses is greater than the discharge time constant $\tau = RC$. Two small pulses not separated in time will appear as one large pulse. To avoid this problem and for noise reduction, a second higher threshold level had been introduced. Now true large pulses cross both thresholds and time-over-high-threshold (HiToT) is a much better measure of the pulse area. It also provides a method of separating small single pulses from everything else.

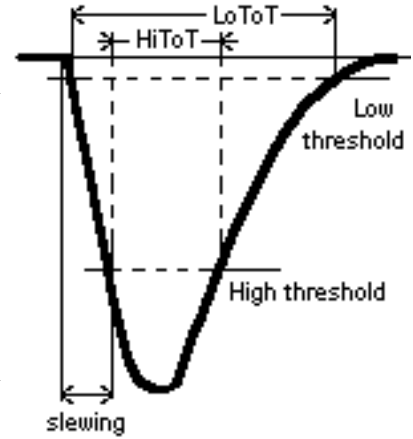

**Figure 1:** Time-over-Threshold concept.

Similarly, the PMT pulse amplitude is related to ToT and the pulse amplitude dependence of TDC measurements can be compensated using ToT.

## 3 Laser calibration system: description

The Milagro calibration system is based on the laser – fiber-optic – diffusing ball concept used in other water-Cherenkov detectors (See, for instance, Becker-Szendy et al., 1995). A computer operated motion controller drives a neutral density filter wheel to attenuate a pulsed nitrogen dye laser beam. The beam is directed to one of the thirty diffusing laser balls through the fiber-optic switch (See Fig 2). Part of the laser beam is sent to a photo-diode. When triggered by the photo-diode, the pulse-delay generator sends a trigger pulse to the data acquisition system. The balls are floating in the pond so that each PMT can "see" more than one light source. Such a redundant setup allows us to calibrate the PMTs and the electronics.

## 4 Timing calibration

Because of finite rise-time of a PMT pulse, its registration time depends on the amplitude of the signal. The corrections were found by studying how TDC outputs vary as a function of PMT-pulse ToT. For different laser pulse intensities, the time of registration ($t_{reg}$) of the PMT response by its TDC with respect to the photo-diode "zero" and ToT were measured. The slewing correcting curve was found by fitting a polynomial to $t_{reg}$ vs $ToT$.

However, since all the time measurements are done with respect to the photo-diode, the slewing curve is artificially shifted by fiber-optic delay, light travel time in water and TDC time offsets. Knowing the locations of the diffusing laser balls and PMTs, the speed of light in water and fiber-optic delay, the TDC offsets can be found.

This procedure has been repeated for both low and high thresholds (LoToT and HiToT) for each PMT-TDC channel. Now, a meaningful interpretation for the TDC outputs exists.

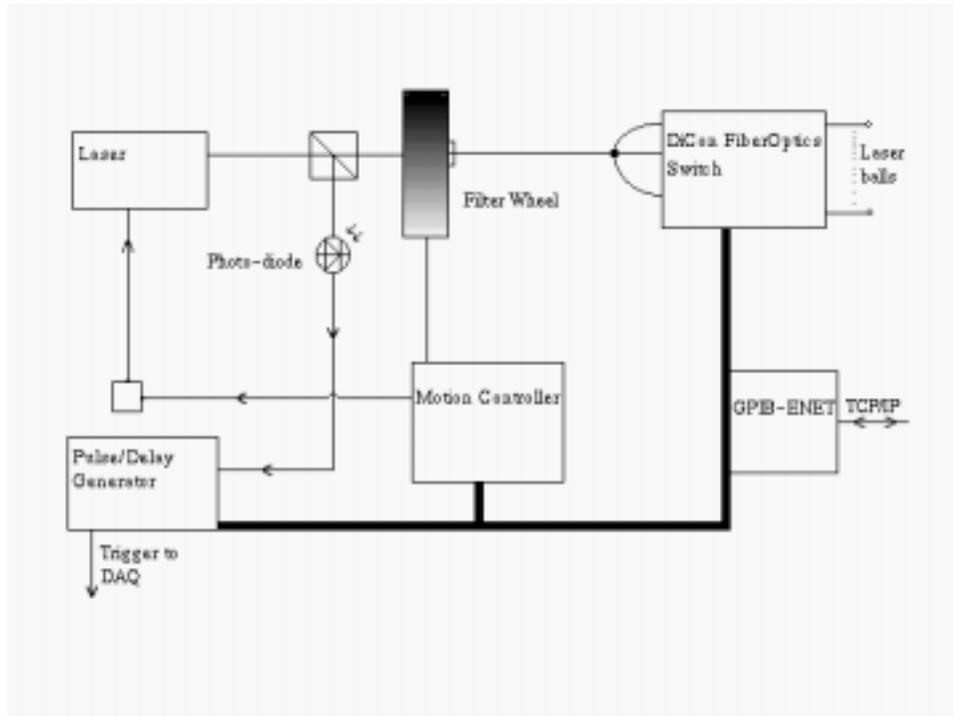

Figure 2: Calibration system setup

## 5 Consistency check

The Milagro calibration system has been designed to allow cross-calibration of the PMTs. This fact was used to check the accuracy of the obtained calibration parameters and disclose some problems.

TDC time offsets obtained for a given PMT from different laser balls should be identical. Thus, the TDC-offset mismatch distribution becomes a useful diagnostic tool. The mean mismatch in offsets over all PMTs from two laser balls gives the fiber-optic delay difference between them. The width of the mismatch distribution is a measure of the offset quality. In fact, if we used a wrong speed of light in water, it would widen the mismatch distribution. Eventually, this allowed us to determine the effective speed of light in the pond water to four decimal places, by comparing measured offsets from a pair of far separated laser balls.

Another use of redundancy was the reconstruction of laser ball coordinates. Direct reconstruction of the coordinates by reversing the procedure in section 4 will yield either a perfect result or will lead to an inconsistency. In either case, it will not give any constructive information. To overcome this difficulty, a method of pairwise correlations was developed to obtain the laser ball locations. The positions of two laser balls s1 and s2 can be obtained by comparing relative time differences from the pair as observed by a PMT. If we define:

$$\tau = t_{s1 \to PMT} - t_{s2 \to PMT}$$

where $t_{s1 \to PMT}$ and $t_{s2 \to PMT}$ are TDC times, then $\tau$ does not depend on the TDC time offset for the particular PMT channel. From geometrical point of view, a given PMT lies on a branch of hyperbola, defined by parameter $\tau$ and with the laser balls s1 and s2 at its foci. Thus, four different PMTs with their corresponding $\tau$'s for the same laser ball pair would define the coordinates of the foci; by using more PMTs the problem becomes overconstrained and yields a best fit. This procedure was used successfully to reconstruct laser ball coordinates for Milagrito and Milagro.

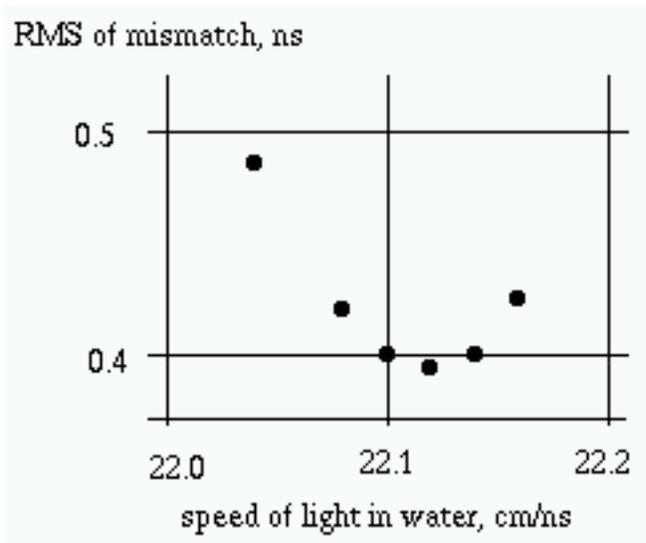

**Figure 3:** Width of the TDC-offset mismatch distribution as a function of speed of light in water.

## 6 Energy calibration

The ToT information was converted to a pulse amplitude scale by moving a set of ADCs to all PMT channels and collecting simultaneous ToT and ADC data. The single photo-electron peak was clearly visible yielding the ToT-to-PE conversion, assuming the ADC outputs are linear in the number of PEs. Alternatively, assuming that the number of registered PEs obeys a Poisson distribution, the occupancy method was used to obtain the ToT-to-PE calibration. Both methods are in reasonable agreement.

Absolute energy calibration measurements will be done using through-going muons. The imaging capabilities of the detector will be exploited in order to find, fit and select well-defined through-going muon tracks. Once the geometry of the track is known, the Cherenkov energy deposit will be estimated and compared against the photo-electron distribution in the event. This was the primary absolute energy calibration method used in the IMB detector (Becker-Szendy et al., 1995).

## 7 Acknowledgment

This research was supported in part by the National Science Foundation, the U. S. Department of Energy Office of High Energy Physics, the U. S. Department of Energy Office of Nuclear Physics, Los Alamos National Laboratory, the University of California, the Institute of Geophysics and Planetary Physics, The Research Corporation, and the California Space Institute.